%% file: main.tex
\documentclass[conference]{IEEEtran}
\IEEEoverridecommandlockouts
\usepackage{cite}
\usepackage[colorlinks=true,linkcolor=black,anchorcolor=black,citecolor=black,filecolor=black,menucolor=black,runcolor=black,urlcolor=magenta]{hyperref}
\usepackage[all=normal, paragraphs, floats, bibnotes, charwidths]{savetrees}
\usepackage{amsmath,amssymb,amsfonts}
\usepackage{algorithmic}
\usepackage{graphicx}
\usepackage{textcomp}
\usepackage{xcolor}
\def\BibTeX{{\rm B\kern-.05em{\sc i\kern-.025em b}\kern-.08em
    T\kern-.1667em\lower.7ex\hbox{E}\kern-.125emX}}

\ifCLASSOPTIONcompsoc
\usepackage[caption=false,font=normalsize,labelfon
t=sf,textfont=sf]{subfig}
\else
\usepackage[caption=false,font=footnotesize]{subfig}
\fi

\newcommand{\mat}[1]{\mathbf{#1}}
\newcommand{\vc}[1]{\mathrm{vec}\left({#1}\right)}

\begin{document}
\bstctlcite{Settings}
\title{Propagation of Measurement and Model Uncertainties through Multiline TRL Calibration}
\author{%
	\IEEEauthorblockN{%
		Ziad Hatab\textsuperscript{12},
		Michael Gadringer\textsuperscript{2}, Wolfgang Bösch\textsuperscript{12}
	}%
	\IEEEauthorblockA{%
		\textsuperscript{1}Christian Doppler Laboratory for Technology Guided Electronic Component Design and Characterization, Graz, Austria\\
		\textsuperscript{2}Institute of Microwave and Photonic Engineering, Graz University of Technology, Graz, Austria\\
		\{z.hatab; michael.gadringer; wbosch\}@tugraz.at
	}%
\thanks{Software implementation available online with additional examples: \url{https://github.com/ZiadHatab/uncertainty-multiline-trl-calibration}}
}%
\maketitle

\begin{abstract}
	In this work, we present a linear uncertainty (LU) propagation treatment of measurement and model uncertainties in multiline thru-reflect-line (TRL) calibration. The proposed method is in accordance with the ISO Guide to the Expression of Uncertainty in Measurement (GUM). We demonstrate that our proposed LU method delivers identical uncertainty compared to numerical simulation based on the Monte Carlo (MC) method, but in a more efficient way. \\
\end{abstract}


\input{Sections/Section1}
\input{Sections/Section2}
\input{Sections/Section3}
\input{Sections/Section4}
\input{Sections/Section5}


\section*{Acknowledgment}
The financial support by the Austrian Federal Ministry for Digital and Economic Affairs and the National Foundation for Research, Technology, and Development is gratefully acknowledged.

\bibliographystyle{IEEEtran}
\bibliography{References/references.bib}

\end{document}

%% file: Sections/Section1.tex
\section{Introduction}
\label{sec:1}
Many vector network analyzer (VNA) self-calibration methods, e.g. (multiline) TRL, allow for partially defined calibration standards. Due to the lack of a direct relationship between the calibration equations and the standards, it becomes challenging to directly assess uncertainties using the conventional LU propagation techniques \cite{jcgm}. Uncertainty propagation methods were proposed for the classical TRL calibration \cite{Hall2018,Zhao2020}. However, for the multiline TRL calibration, there is less literature available. Currently, the usual approach for uncertainty evaluation in multiline TRL is to use the MC method \cite{Luan2020}, which can be time-consuming and inefficient. The difficulty in applying the LU method in multiline TRL calibration stems from the mathematical limitation of the original algorithm \cite{Marks1991}. The multiline TRL calibration was formulated by perturbing the solutions of TRL pairs, where a common line is selected as a reference among all line pairs. Furthermore, the implementation of multiline TRL in \cite{Marks1991} requires the numerical computation of an inverse covariance matrix, which can lead to unstable results.

In a recent publication \cite{Hatab2022}, we presented a new mathematical formulation of the multiline TRL calibration problem. We formulated the calibration problem into solving a single $4\times4$ eigenvalue problem, regardless of how many line standards were used. We showed that our method is more reliable than \cite{Marks1991}. Moreover, our technique's mathematical simplicity allows us to propagate different types of uncertainties using the LU method. In the following sections, we discuss how to treat different types of uncertainties and demonstrate that our method delivers identical results as the MC method, but in a much more efficient way.


%% file: Sections/Section2.tex
\section{Multiline TRL Equations}
\label{sec:2}

This section summarizes the basic equations used to solve the multiline TRL problem, as discussed in \cite{Hatab2022}. In contrast to many authors, we write the 7-term error box model for the measured scattering transfer (T-) parameters of the $i$th line standard based on the Kronecker product formulation that we presented in \cite{Hatab2022}:
\begin{equation}
	\vc{\mat{M}_i} = k\underbrace{(\mat{B}^T\otimes\mat{A})}_{\mat{X}}\vc{\mat{L}_i},
	\label{eq:2.1}
\end{equation}
where $\mat{A}$ and $\mat{B}$ are the left and right error boxes holding the first six error terms, while $k$ is the $7$th error term. We showed in \cite{Hatab2022} that the calibration matrix $\mat{X}$ can be determined by solving the following weighted eigenvalue problem:
\begin{equation}
	\mat{M}\mat{W}\mat{D}^{-1}\mat{M}^T\mat{P}\mat{Q} = \mat{X}\begin{bmatrix}
	-\lambda & 0 & 0 & 0\\
	0 & 0 & 0 & 0\\
	0 & 0 & 0 & 0\\
	0 & 0 & 0 & \lambda
\end{bmatrix}\mat{X}^{-1}
	\label{eq:2.2}
\end{equation}

Without going into detail, the most relevant aspect of (\ref{eq:2.2}) is that the left-hand side consists purely of the measurement of the line standards, which are optimally weighted by the matrix $\mat{W}$. The right-hand side of (\ref{eq:2.2}) describes the combined model, which is an eigenvalue problem in $\mat{X}$. A detailed discussion of the matrices in (\ref{eq:2.2}) is given in \cite{Hatab2022}. After solving for $\mat{X}$ from (\ref{eq:2.2}), we must denormalize the solution uniquely. This denormalization is achieved by measuring the reflect and the thru standards while simultaneously solving for the $7$th error term $k$.


%% file: Sections/Section3.tex
\section{Uncertainty Treatment}
\label{sec:3}
To assess uncertainties in our multiline TRL implementation, we categorize the uncertainties into three types. 
\subsection{Measurement uncertainty}
These are uncertainties that arise from the measurement equipment itself. For example, this could be additive noise or phase noise. The corresponding covariance matrix of the measured S-parameters can be determined from the wave quantities through transformation using the appropriate Jacobian matrix. Then, knowing the covariance matrix, we can use the LU method to propagate the uncertainties directly.

\subsection{Forward model uncertainty}
Certain parameters, such as the length of the line standards, are related directly to the solution of the calibration coefficients. We can propagate the uncertainty of such parameters directly using the LU method. Even for the unknown, symmetric reflect standard, we can still propagate its uncertainty by keeping it as an independent variable and only substituting its estimated value in the corresponding Jacobian matrix.

\subsection{Inverse model uncertainty}
Every other type of uncertainty that is implicit in the solution model should be addressed by inverse uncertainty quantification. For example, to construct the eigenvalue problem in (\ref{eq:2.2}), we require the assumption that all line standards share an identical characteristic impedance. However, because of random variations, such assumptions are prone to uncertainty. Therefore, a direct LU propagation for impedance mismatch is not possible. The solution is to inversely update the covariance matrix of the measurement of each line standard. Generally, the covariance matrix of the model-based measurement is given by the superposition of the covariance of the measurement and the model \cite{Wu2019}:
\begin{equation}
	\mat{\Sigma} = \mat{\Sigma}_\mathrm{meas} + \mat{\Sigma}_\mathrm{model}
	\label{eq:3.1}
\end{equation}

Since multiline TRL is an averaging algorithm, we can first determine the calibration coefficients without uncertainty treatment and then use the values obtained to derive $\mat{\Sigma}_\mathrm{model}$. After that, we update the covariance matrix of each measurement and perform the LU propagation as usual.

%% file: Sections/Section4.tex
\section{Numerical Validation}
\label{sec:4}
We performed an MC simulation to validate that our LU method can handle the various types of uncertainty simultaneously. For this purpose, we used numerically generated multiline TRL data \cite{skrf}. In this way, we have complete control of each aspect of the standards. To emulate an actual measurement scenario, we simulated a set of coplaner waveguide (CPW) lines connected to coaxial cables. We performed 1000 MC trials, where we introduced randomness in the embedded standards (i.e., additive noise), the length of the standards, and the non-embedded reflect standard. A mismatch among the line standards is also simulated by introducing randomness in the permittivity of the dielectric substrate. We simulated a lossless, symmetrical, and equal reflection and transmission network as a device under test (DUT) to verify the calibration. For our convenience, we used the Metas.UncLib package \cite{Zeier2012} for dynamic computation of all needed derivatives. However, it is also possible to use other methods, such as eigenvector perturbation techniques or pre-computing all derivative expressions and storing them in a lookup table.

\begin{figure}[th!]
	\centering
	\includegraphics[width=0.98\linewidth]{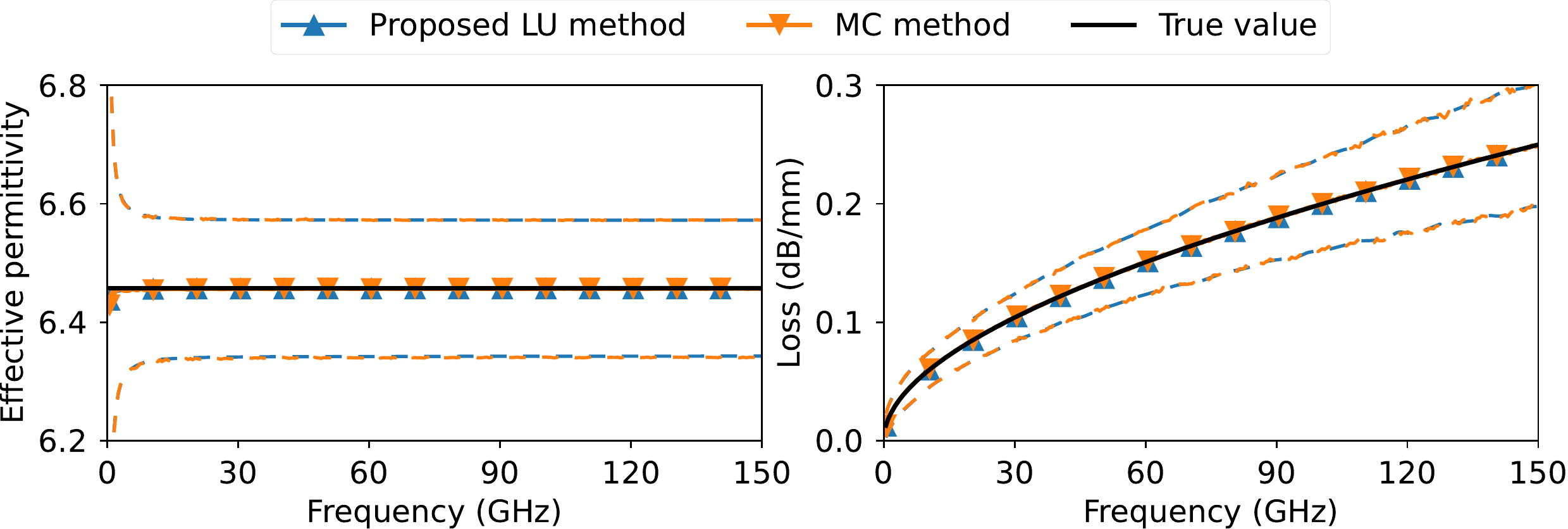}
	\caption{Effective permittivity and loss per-unit-length of the simulated CPW lines.}
	\label{fig:4.1}
\end{figure}

\begin{figure}[th!]
	\centering
	\includegraphics[width=0.98\linewidth]{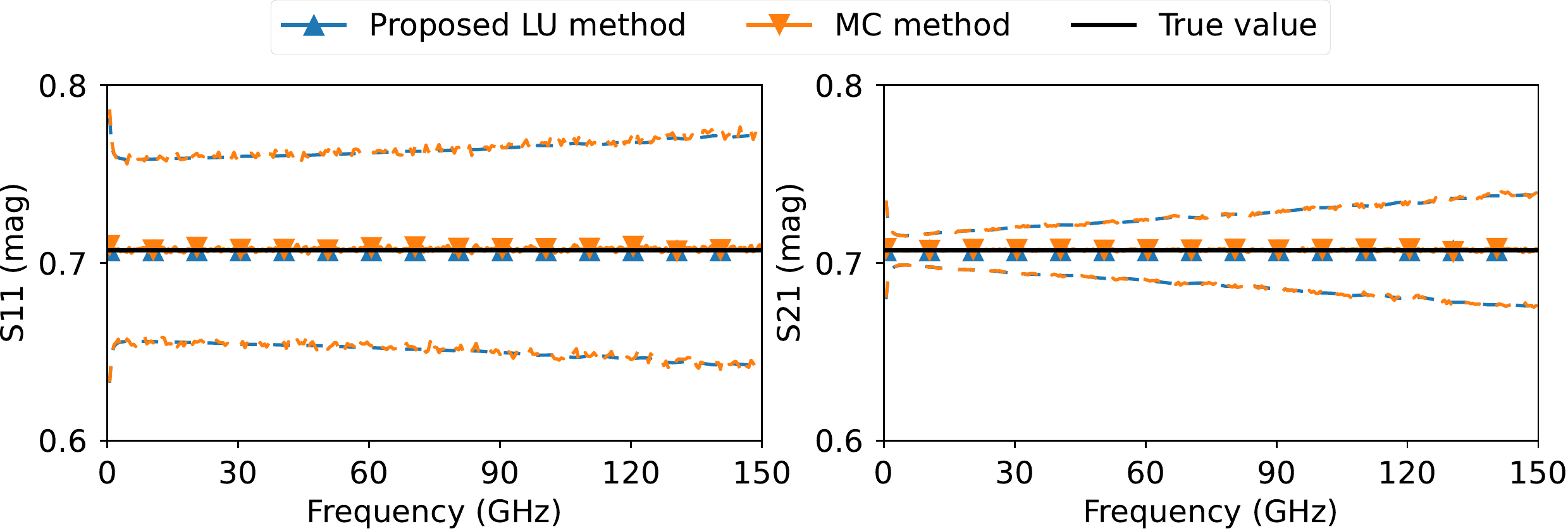}
	\caption{Magnitude of $S_{11}$ and $S_{21}$ of the calibrated DUT.}
	\label{fig:4.2}
\end{figure}

The result of the numerical simulation is presented in Figs. \ref{fig:4.1} and \ref{fig:4.2}. We can distinctively see that our LU method and the MC approach delivered identical uncertainties. The time it took to run 1000 MC trials was around 18\,min, whereas our LU method (with Metas.UncLib) took less than one minute to finish.


%% file: Sections/Section5.tex
\section{Conclusion}
\label{sec:5}
We demonstrated how to treat different types of uncertainties, including non-trivial cases, and how to propagate them through multiline TRL calibration using the LU propagation method according to GUM. Furthermore, we showed that our LU method delivered identical uncertainty bounds compared to the MC simulations. In contrast to the MC method, the LU method required a significantly shorter computation time.